\documentstyle[12pt,supercite]{article}

\newcommand{\BF}[1]{\mbox{\boldmath $#1$}}

\def\abstracts#1#2#3{{
        \centering{\begin{minipage}{4.62in}\baselineskip=13pt
        \small
        \centerline{\bf Abstract}
        \vspace*{0.2cm}                
        \parindent=0pt #1\par
        \parindent=18pt #2\par
        \parindent=15pt #3
        \end{minipage} }\par}}
\renewcommand{\thefootnote}{\fnsymbol{footnote}}
\begin{document}
\vspace*{-2cm}
\hfill \parbox{4cm}{ FUB-HEP 10/93 \\
              \today }\\
\vspace*{2cm}
\centerline{\LARGE \bf Monte Carlo Study of Topological
Defects}\\[0.3cm]
\centerline{\LARGE \bf in the 3D Heisenberg
Model\footnotemark}\\[0.4cm]
\footnotetext{\noindent Work supported in part by Deutsche
Forschungsgemeinschaft under grant Kl256.}
\addtocounter{footnote}{-1}
\renewcommand{\thefootnote}{\arabic{footnote}}
\vspace*{0.2cm}
\centerline{\large {\em Christian Holm\/}$^1$ and
                   {\em Wolfhard Janke\/}$^2$}\\[0.4cm]
\centerline{\large    $^1$ {\small Institut f\"{u}r Theoretische
Physik,
                      Freie Universit\"{a}t Berlin}}
\centerline{    {\small Arnimallee 14, 1000 Berlin 33,
Germany}}\\[0.2cm]
\centerline{\large    $^2$ {\small Institut f\"ur Physik,
                      Johannes Gutenberg-Universit\"at Mainz}}
\centerline{    {\small Staudinger Weg 7, 6500 Mainz 1, Germany
}}\\[0.5cm]
\abstracts{}{
We use single-cluster Monte Carlo simulations to study the
role of topological defects in the three-dimensional classical
Heisenberg
model on simple cubic lattices of size up to $80^3$. By applying
reweighting
techniques to time series generated in the vicinity of the
approximate infinite
volume transition point
$K_c$, we obtain clear evidence that the temperature derivative
of the average defect density $d\langle n \rangle/dT$ behaves
qualitatively
like the specific heat, i.e., both observables are finite in the
infinite
volume limit. This is in contrast to results by
Lau and Dasgupta [{\em Phys. Rev.\/} {\bf B39} (1989) 7212]
who extrapolated a divergent behavior of $d\langle n \rangle/dT$ at
$K_c$
from simulations on lattices of size up to $16^3$. We obtain weak
evidence that $d\langle n \rangle/dT$ scales with the same critical
exponent as the specific heat.
As a byproduct of our simulations, we obtain a very accurate
estimate for the
ratio
$\alpha/\nu$ of the
 specific-heat
exponent with the correlation-length exponent from a finite-size
scaling analysis of
the energy.}{}
\vspace*{1.5cm}
PACS number(s): 75.10.Hk, 02.70.Lq, 75.40.Mg, 05.70.Fh
\thispagestyle{empty}
\newpage
\pagenumbering{arabic}
%
                     \section{Introduction}
%
It is well known that
topological defects can play an important role in phase transitions
\cite{halp,klei}. Extensively studied examples of systems with
pointlike
defects are the two-dimensional (2D)
XY model \cite{kos} and defect models for 2D melting \cite{klei,wj}.
Recently
Lau and Dasgupta (LD) \cite{lau} have used Monte Carlo (MC)
simulations
to study the role of topological defects in the three-dimensional
(3D)
classical Heisenberg model, where the defects are also point-like
objects with
a binding energy that increases linearly with the separation
\cite{os81}.
Motivated
by the importance of vortex points in the 2D XY model, LD tried to
set up a
similar
pictorial description of the phase transition in the 3D Heisenberg
model.
Analyzing their simulations on simple cubic (sc) lattices of size
$V=L^3$ with
$L=8,12$ and $16$, LD claimed that the temperature derivative of the
average
defect density, $\langle n \rangle$, diverges at the critical
temperature
$T_c$ like $d\langle n \rangle/dT \sim t^{-\psi}$, $t =
|T-T_c|/T_c$, with
an exponent $\psi \approx 0.65$.
They further speculated that $\psi=1-\beta$, where $\beta \approx
0.36$ is the
critical exponent of the magnetization, and then argued that
$\langle n \rangle$
should behave like a ``disorder'' parameter.

At first sight, the existence of such a strong divergence of
$d\langle n \rangle/dT$
seems unlikely, because the
definition of defects is quasi-local. It is therefore more likely
\cite{sokal-private} that
$\langle n
\rangle$ should qualitatively behave like the energy and
$d\langle n \rangle/dT$ like
the specific heat, which is a finite quantity for the 3D Heisenberg
model.

Using standard finite-size scaling (FSS) arguments
we hence expect to see on finite lattices either
\begin{equation}
d\langle n \rangle/dT = L^{\psi/\nu} f(x)
\label{eq:LD}
\end{equation}
or, if the second argument holds true,
\begin{equation}
d\langle n \rangle/dT = const + L^{\alpha/\nu} g(x),
\label{eq:sokal}
\end{equation}
where $\nu \approx 0.7$ and $\alpha \approx -0.1$ are the
correlation length
and specific heat exponents, respectively, $x=tL^{1/\nu}$ is the
finite-size scaling variable, and $f(x),g(x)$ are scaling functions.
At fixed $x$, Ansatz (\ref{eq:LD}) predicts an approximate linear
divergence in $L$, $d \langle n \rangle/dT \sim L^{\psi/\nu} \approx
L$.
On the other hand, because $\alpha$ is {\it negative} for the 3D
Heisenberg
model, Ansatz (\ref{eq:sokal}) predicts a constant asymptotic value,
$d \langle n \rangle /dT \sim const$.
For sufficiently large $L$ a clear distinction between (\ref{eq:LD})
and
(\ref{eq:sokal}) should hence be observable.

To decide between the two alternatives we have performed MC
simulations on
large lattices of size up to $L=80$, employing the single-cluster
update
algorithm \cite{wolff} and reweighting techniques \cite{r20}.
As a result we find unambiguous support
for the second alternative, $d \langle n \rangle /dT \sim const$,
as $L \longrightarrow \infty$.

As a byproduct of our simulations we numerically extrapolate
$\alpha/\nu$ from a FSS of the energy in
close vicinity of the critical
coupling $K_c$ with a much higher accuracy than obtained
in recent
high-precision MC studies \cite{landau1,wir}.

%
            \section{The simulation}
%

The partition function of the Heisenberg model is given by
\begin{equation}
Z = \prod_{i} \left[ \int
\frac{d\Omega_i}{4\pi} \right] \exp(-KE);
{}~~E = \sum_{\langle i,j \rangle}
(1-\vec{s}_i \cdot \vec{s}_j).
\label{eq:1}
\end{equation}
where $K \equiv J/k_BT$ is the (reduced) inverse temperature,
 $\vec{s}_i$ are three-dimensional unit vectors at the sites $i$ of
a
sc lattice, and $\langle i,j \rangle$ denotes nearest-neighbor
pairs.
Using the single-cluster update algorithm \cite{wolff} we have
simulated the partition function (\ref{eq:1}) for lattices of size
$V=L^3$ with $L$=8,12,16,20, 24, 32, 40, 48, 56, 64, 72,
80 and periodic boundary conditions.
Our main emphasis was on the defect density $n = \sum q^2n_{|q|}$,
where
 $n_1, n_2, \dots$  are defect densities
of charge $q=\pm 1, \pm 2, \dots$. To locate these charges we
followed the definition of Berg and L\"uscher \cite{berg} according
to which
the charge $q_{i^*}$ at the dual lattice site $i^*$ is given
by
\begin{equation}
q_{i^*} = \frac{1}{4\pi} \sum_{i=1}^{12} A_i,
\label{eq:2a}
\end{equation}
with
\begin{equation}
\cos(\frac{1}{2} A_i) = \frac{1 + \vec{s}_1 \cdot \vec{s}_2
                              + \vec{s}_2 \cdot \vec{s}_3
                              + \vec{s}_3 \cdot \vec{s}_1}
                           {\sqrt{2(1+\vec{s}_1 \cdot \vec{s}_2)
                                   (1+\vec{s}_2 \cdot \vec{s}_3)
                                   (1+\vec{s}_3 \cdot \vec{s}_1)}},
\label{eq:2b}
\end{equation}
and the sign of $A_i$ is determined by ${\rm sign} A_i = {\rm sign}
\left( \vec{s}_1 \cdot (\vec{s}_2 \times \vec{s}_3)\right)$.
For a sc lattice the distinction between $i$ and $i^*$ is
inessential, since the
difference is only a uniform translation along the
space diagonal. The sum
$\sum_{i=1}^{12}$ in (\ref{eq:2a}) refers to the 12 triangles that
can
be formed on the faces of the cube enclosing $q_{i^*}$. In
(\ref{eq:2b})
the spins
at the corners of these triangles are numbered in a
counter-clockwise sense
relative to the outward pointing normal. For the orientation of
the diagonals on each face of the cube we used the convention that
they run
from the lattice point
$i$ to $i + \BF{e}_i + \BF{e}_j$, and
from $i + \BF{e}_i$ to $i + \BF{e}_i + \BF{e}_j +
\BF{e}_k$,
where $\BF{e}_i, i=1,2,3$ denote unit-vectors in the three
coordinate directions. This choice is obviously not unique, but we
have checked
that other conventions give on the average the same charges within
the error
bars. From the definition of $q_{i^*}$ it is clear, that a trivial
upper bound on the
magnitude of the
lattice topological charge is $q_{i^*} \le 5$. In our runs the
highest
topological
charge observed was three, which occured on the order of $10^{-7}$
per site and
measurement, see Table~1. The likelihood of the appearence of the
higher charges
was probably too
small for them to occur during our run times.

All runs were performed close to the approximate infinite volume
transition
point
$K_c=0.6930$, as determined in recent MC studies
\cite{landau1,wir,landau2} of
this
model. Since we wanted to have reference data we performed our
simulation at the
same coupling $K_0 = 0.6929$ as in our study of ref. \cite{wir},
which is
close enough to $K_c$ to allow safe reweighting of our data to this
value of
$K_c$.
Because the computation of $q_{i^*}$ is quite complex and thus time
consuming,
we have
performed many cluster update steps between measurements, adjusted
in such a way
that the (integrated) autocorrelation time of the charge density
measurements is
around $\tau_n \approx 1-2$. Since it turned out that the
(integrated)
autocorrelation
times $\tau_n$ and $\tau_\chi$ of the charge density and the
magnetic
susceptibility are roughly equal, we were able to guess the required
measurement
interval by extrapolating our previous results for $\tau_\chi$
\cite{wir} to
larger
lattice sizes $L$. The measurement statistics are given in Table~1.
While the
statistics is comparable to that of our previous studies \cite{wir},
and much
better
than that of LD, we note, that our investigated lattices have much
 larger linear size up to
$L$=80 as compared to $L$=48 in our previous work, and compared to
the largest
size
$L=16$
of LD.  For each run we recorded the time series of the energy
density $e=E/V$,
the
magnetization density $m=| \sum _i \vec {s}_i|/V$, and the charge
densities
$n_{|q|}$.
The resulting averages $\langle e \rangle$,  $\langle n \rangle$,
and $\langle
n_{|q|}
\rangle$ can be found in Table~1.

To
compute the specific heat $C = d \langle e \rangle /dT$, the thermal
expansion
coefficient  $C_q=Td \langle n \rangle /dT$, and the topological
susceptibility
$\chi_q=d \langle n \rangle /d\mu$,  where $\mu$ is  the ``field''
in a fugacity
term
$\mu \sum_{i^*} q_{i^*}^2$ which one can imagine adding to the
energy in
(\ref{eq:1}),
and $n$ is defined as $n= n_1 + 4n_2 + 9n_3 + \dots$, we used the
relations
\begin{eqnarray} C &=& VK^2 ( \langle e^2 \rangle - \langle e
\rangle^2 ) = V
K^2
      \langle e;e \rangle, \label{eq:3a} \\
C_q &=& V K ( \langle e n \rangle - \langle e \rangle
             \langle n \rangle) = V K \langle e;n \rangle,
\label{eq:3b}\\
\chi_q &=& V ( \langle n^2 \rangle - \langle n \rangle^2 ) = V
      \langle n;n \rangle. \label{eq:3c}
\end{eqnarray}
To obtain results for the various observables ${\cal O}$ at $K$
values in
an interval around the simulation point $K_0 = 0.6929$, we applied
the
reweighting method \cite{r20}.
Since we recorded the time series this amounts to computing
\begin{equation}
\langle {\cal O} \rangle |_{K} =
\frac{\langle {\cal O} e^{-\Delta K E} \rangle |_{K_0}}
     {\langle e^{-\Delta K E} \rangle |_{K_0}},
\label{eq:4}
\end{equation}
with $\Delta K = K - K_0$. To obtain errors we devided each run into
20 blocks and used
standard Jackknife errors \cite{r27}.

The results for the quantities in (6) - (8) at $K_c = 0.6930$
are collected in Table~2. Also given are the eigenvalues
$\lambda_1,\lambda_2$
of the $2 \times 2$ covariance matrix $M$ of $e$ and $n$ with
elements
$M_{11} = V K^2 \langle e;e \rangle$,
$M_{12}=M_{21}=V K \langle e;n \rangle$,
$M_{22} = V \langle n;n \rangle$.

%
   \section{Results}
%
Applying (\ref{eq:4}) we have determined the
temperature dependence of the quantities in
(\ref{eq:3a})-(\ref{eq:3c}). For small lattices, $C_q$ has its peak
location at temperatures larger than $T_c$, in contrast to $C$,
which
peaks at temperature values smaller than $T_c$.
With increasing lattice size, however, we observe a strong
correlation between
$C_q$ and $C$, that is, both quantities develop a smooth peak at
roughly the
same temperature ($T<T_c$), see Fig.~1. In contrast to $C$ the peak
locations of
$C_q$ scale non-monotonically, with a crossover at $L \approx 20$.

We focussed first on the scaling behavior of $C_q$ at our previous
estimate of
the
critical coupling $K_c = 0.6930$, obtained from the crossings of the
Binder
parameter
$U = 1 - \langle m^4\rangle / 3 \langle m^2\rangle ^2$ \cite{wir}.
Our new data for
$U$ on the large lattices confirmed the constancy of our previous
result of $U^* = 0.6217(8)$ and hence our
estimate for $K_c$. We checked
first a scaling Ansatz  for $C_q$ of the form \begin{equation} C_q =
C_{q}^{\rm reg} - a_0L^{\alpha'/\nu}, \label{eq:c_q}
\end{equation}
where $C_q^{\rm reg}$ is a regular background term which is assumed
to be
independent of lattice size
\cite{privman}. Note that this Ansatz covers both scaling hypotheses
(\ref{eq:LD}) and (\ref{eq:sokal}).
The resulting fit shown in Fig.~2(a) yields $\alpha'/\nu =
-0.401(61)$, $C_{q}^{\rm reg} = 1.50(8)$, and $a_0 =1.82(6)$,
with a quality factor \cite{r24} $Q= 0.30$.
The good
quality of the fit basically rules out the divergence predicted by
the Ansatz (\ref{eq:LD}) of LD,
and
strongly favours (\ref{eq:sokal}), which predicts a finite
asymptotic
value for $C_q$. Only if one assumes that the FSS
behavior sets
in at extremely large lattice sizes, one could still attain an
assertion
of the
form
(\ref{eq:LD}), but with the consequence of an extremely small
exponent
$\psi$.
We also tried to reproduce the exponent $\psi \approx 0.65$ of LD,
by selecting
only their lattices sizes, and fitting a straight line to our first
3
data
points. But even then we obtain a much smaller value of
$\psi /\nu \approx 0.36(3)$, leading to $\psi \approx 0.25(3)$.
We think, this discrepancy to the result of LD is
partly due to our higher statistics and partly due to the fact that
we
obtained
$C_q$
through  a thermodynamic derivation, which normally gives better
results than the numerical
differentiation used by LD.

Because the Ansatz (\ref{eq:sokal}), which was based on the
assumption
that $\langle n \rangle$ should behave like the energy, fits so
well,
one can ask,
if $\alpha'$ is equal to the specific-heat exponent $\alpha$.
Using our earlier MC
result \cite{wir} of  $\nu = 0.704(6)$, we get a value of
$\alpha' = -0.282(46)$, which does, on the first glance, not
strongly support
this conjecture.
The best field theoretical estimates are
$\nu = 0.705(3)$, $\alpha = -0.115(9)$, and $\alpha/\nu =
-0.163(12)$
(resummed perturbation series \cite{zi80}), while our earlier MC
study
\cite{wir} yielded $\nu = 0.704(6)$, $\alpha = -0.112(18)$, and
$\alpha/\nu
= -0.159(24)$. However, the accuracy of the
values of $\alpha$ is somewhat misleading,
because they were obtained from hyperscaling, $\alpha = 2 - 3\nu$.
The
directly measured values have much larger error bars, for example
$\alpha/\nu = -0.30(6)$ \cite{landau1} and $\alpha/\nu = -0.33(22)$
\cite{wir}.

To compare $\alpha'$ directly
with the
measured specific-heat exponent of the present MC simulation, we
fitted $C$ to
\begin{equation} C = C^{\rm reg} - b_0L^{\alpha/\nu},
\label{eq:c}
\end{equation}
with a constant background term $C^{\rm reg}$ \cite{privman}.
The resulting fit in Fig.~2(b) yields $\alpha/\nu =-0.225(80)$,
$C^{\rm reg} = 4.8(7)$,
and $b_0 = 4.1(5)$ with $Q=0.55$,
leading to $\alpha = -0.158(59)$. These values are in very good
agreement with
the
hyperscaling prediction, but noteworthy is also the tendency for the
values to
come out too large.

Of course, a fit of a divergent quantity, like the
first derivative $dC_q/dT$, for example, is in principle numerically
much easier
to
handle.
We tried to do this for $dC_q/dT$ and $dC/dT$ at $K_c$, and observed
the expected divergent scaling behavior, but unfortunately the
statistical
errors of the third cumulants involved turned out to be much
too large to allow for meaningful fits.

Other estimates for $\alpha$ and $\alpha'$ can be obtained by means
of fits
of $\langle e \rangle$ and $ \langle n \rangle$, which
again look qualitatively very alike. According to (\ref{eq:c}),
we should have on periodic lattices a scaling behavior of the
energy density $\langle e \rangle$ of the form \cite{privman,mon}
\begin{equation}
\langle e \rangle = \langle e \rangle ^{\rm reg} -
d_0L^{(\alpha - 1)/\nu}.
\label{eq:e}
\end{equation}
and because of (\ref{eq:c_q}) the topological charge density
$\langle n \rangle$
should then accordingly scale like
\begin{equation}
\langle n \rangle = \langle n \rangle ^{\rm reg}  -
c_0L^{(\alpha' -1)/\nu}.
\label{eq:n}
\end{equation}
Fits of these quantities at $K_c = 0.6930$, shown in Fig.~3, yield
$(\alpha'  -
1)/\nu =-1.547(15)$, $\langle n \rangle ^{\rm reg} = 0.1074(1)$, and
$c_0 =
0.42(2)$,
 with $Q = 0.30$, and $(\alpha - 1)/\nu =-1.586(19)$,
$\langle e \rangle ^{\rm reg} = 2.0106(1)$,
and $d_0 = 1.68(8)$,
with $Q=0.25$. This results in $\alpha'/\nu =-0.127(27)$, $\alpha' =
-0.089(20)$,  $\alpha/\nu =-0.166(31)$, and $\alpha = -0.117(23)$.
The results for $\alpha$ and $\alpha/\nu$ are in excellent agreement
with the
hyperscaling prediction, and have not been directly measured before
with such a
high precision. We attribute this to our large lattice sizes used,
but also to
the fact that we used fits of $\langle e \rangle$ instead of $C$.
The results
for
$\alpha'$ and $\alpha'/\nu$ are now lower than those obtained in
(\ref{eq:c_q}),
but now they are almost consistent with the values for $\alpha$
and $\alpha/\nu$. Still it is a little bit puzzling that both
estimates
for the exponent
$\alpha'/\nu$
obtained from  the fits (\ref{eq:c_q}) and (\ref{eq:n}) do not agree
in their
respective
error range. We attribute this partly to the unknown  FSS behavior
of
the regular background term $\langle n \rangle ^{\rm reg}$, and
partly to the
fact that the
statistical errors of the three parameter fits should be taken with
great care.

We further looked at the scaling behavior of $\chi_q$, defined in
eq.(\ref{eq:3c}).
A first look at the plots suggests to try again a scaling Ansatz of
the form
\begin{equation}
\chi_q = \chi_{q} ^{\rm reg} - e_0L^{\alpha''/\nu}
\label{eq:chi_q}.
\end{equation}
{}From a three-parameter fit we obtain $\alpha''/\nu =-0.554(57)$,
$\chi_{q} ^{\rm reg} = 0.67(2)$, and $e_0 = 0.95(6)$ with $Q= 0.41$,
leading to $\alpha'' = -0.390(44)$. This time it seems already very
unlikely,
that $\alpha''$ is equal to the specific-heat exponent. However, if
one discards
the two lowest $L$ values from the fit, one observes a clear trend
towards
a lower $\alpha''$-value, but with the drawback of increased error
bars
and no improvement in $\chi^2$/dof (per degree of freedom).

We also checked
in all other fits if there were corrections to FSS, by discarding
successively the
data points for $L=8$ and $L=12$. We observed in all quantities a
trend
to
the value of $\alpha/\nu$ predicted by hyperscaling, but at the
price
of much
larger error bars. Also the $\chi^2$/dof did
not improve.
We further checked for confluent corrections \cite{wegner}, by
including a
term of
the form
$a_1L^{-\omega}$, with $\omega = \Delta/\nu$ fixed at the literature
value 0.78
\cite{zi80}. But again the fits were too unstable to give conclusive
results.

We also tested if our results depended strongly on the choice of
$K_c$,
by repeating the fits of all quantities at $K_c \pm 0.0002$. The
resulting
parameters were
 always  consistent with the values at $K_c$ in the
one-$\sigma$ range.

To get a clearer picture we further looked at the scaling behavior
of the
eigenvalues of the covariance matrix $M$ of $e$ and $n$, defined by
equations (\ref{eq:3a}) - (\ref{eq:3c}), which give two uncorrelated
observables
$\lambda_1$ and $\lambda_2$. Again we used a scaling Ansatz of the
form
\begin{equation}
\lambda_i = \lambda_{i} ^{\rm reg} - a_iL^{\alpha_i/\nu}
\label{eq:lambda}.
\end{equation}
As results we obtain $\alpha_1/\nu = -0.273(73)$,
$\lambda_{1} ^{\rm reg} = 5.1(5)$, and $a_1 = 4.7(2)$,
 with $Q=0.49$ and
$\alpha_2/\nu = -1.45(42)$, $\lambda_{2} ^{\rm reg} = 0.1307(8)$,
and $a_2 =
0.2(2)$,
 with $Q=0.60$, leading to $\alpha_1 =-0.192(54)$ and
$\alpha_2 = -1.02(31)$. This suggests $\alpha_1 \approx \alpha$ and
$\alpha_2
\approx \alpha -1$. The quality of the fits can be inspected in
Fig.~4. The
existence of an uncorrelated observable which scales with an
exponent different
from $\alpha$ suggests that there is possibly a new scaling field,
and that
$C_q$ and $\chi_q$ see remnants of this scaling field in their FSS
behavior.
This becomes particularly clear if one remembers that $\lambda_1 +
\lambda_2 = C + \chi_q$. Therefore at least $\chi_q$ should see
something of the exponent $\alpha_2 \approx \alpha -1$.
Another alternative would be, that $C_q$ and $\chi_q$ do not scale
directly
with $\alpha/\nu$ but with some rational multiple of $\alpha/\nu$.
As long
as there is no satisfactory theory of the scaling of topological
quantities,
however, one cannot decide between these alternatives.

%
                       \section{Concluding remarks}
%
We have shown that in the three-dimensional classical Heisenberg
model
the topological defect density $\langle n \rangle$ and its
temperature
derivative $C_q$ behave qualitatively like the energy
$\langle e \rangle$ and its temperature derivative $C$. Especially,
we can
reject
the conjecture of LD that $C_q$ diverges with a new critical
exponent $\psi$.

Rather, our simulations indicate that $C_q$ behaves also
quantitatively
like the specific heat, i.e. scales like (\ref{eq:sokal}).
We obtain weak evidence that
asymptotically for large $L$ the scaling of $C_q$ is governed by
the specific-heat critical exponent $\alpha$. Still, it cannot be
ruled out that the scaling of $C_q$ involves also a new exponent
belonging to a
new scaling field. For the topological susceptibility $\chi_q$ we
find that it
also remains finite, and that it can be fitted with an Ansatz of the
form
(\ref{eq:sokal}) as well, but that its scaling exponent must be some
multiple of
$\alpha$, be the representative of a new scaling field or be a
mixture of
both. Our fits of the eigenvalues $\lambda_i$ of the covariance
matrix seem
to indicate
that $C_q$ and $\chi_q$ are a mixture of a part which scales with
$\alpha$, and
a part which
scales according to $\alpha - 1$.

Finally, the present fits of the specific heat at $K_c$ yielded a
value of
$\alpha$ of better
accuracy and in better agreement with the hyperscaling value than
fits of the
specific-heat maxima as used in previous works \cite{landau1,wir},
which we
attribute to our large lattice sizes, the larger number of available
data
points,
 and to the fact, that our data and fit
was done extremly close to the critical temperature.
Moreover, by
fitting the {\it energy} at $K_c$ to eq. (\ref{eq:e}), we obtained
an estimate
for
$\alpha/\nu$ with a precision so far unpreceeded by direct numerical
MC
simulations
and in accuracy comparable to hyperscaling predictions.
%
                 \section*{Acknowledgement}
%
We are indepted to
A.D. Sokal for discussions that initiated this study.
W.J. thanks the DFG for a Heisenberg fellowship.\\
The numerical simulations were performed on the CRAY X-MP and Y-MP
of
the Konrad-Zuse Zentrum f\"ur Informationstechnik Berlin (ZIB),
and the CRAY X-MP at the Rechenzentrum der Universit\"at Kiel.
We thank all institutions for their generous support.
%
     
%
%
\newpage
%
%
\begin{table}              

{\Large\bf Tables}\\[1cm]
 \begin{center}
  \begin{tabular}{|r|r|r|r|c|c|c|c|c|}
\hline
\multicolumn{1}{|r|} {$L$}          &
\multicolumn{1}{c|}{$N_0$}  &
\multicolumn{1}{c|}{$N_{\rm meas}$} &
\multicolumn{1}{c|}{$\tau_n$}       &
\multicolumn{1}{c|}{$\langle e \rangle $}            &
\multicolumn{1}{c|}{$\langle n \rangle \times 10$}   &
\multicolumn{1}{c|}{$\langle n_1 \rangle \times 10$}   &
\multicolumn{1}{c|}{$\langle n_2 \rangle\times 10^4$}  &
\multicolumn{1}{c|}{$\langle n_3 \rangle\times 10^8$}  \\
\hline
 8 & 17 &50178  & 1.2 &1.9487(9) &0.9054(18)& .88090 &6.13 &7.78\\
12 & 20 &159575 & 1.6 &1.9786(3) &0.9845(7) & .95689 &6.90 &5.80 \\
16 & 40 &64368  & 1.2 &1.9905(3) &1.0170(9) & .98838 &7.16 &7.21 \\
20 & 50 &27670  & 1.3 &1.9968(3) &1.0347(6) & 1.0053 &7.35 &5.42 \\
24 & 50 &20000  & 1.5 &1.9998(2) &1.0431(6) & 1.0136 &7.38 &9.04 \\
32 & 68 &25403  & 1.5 &2.0045(2) &1.0561(4) & 1.0260 &7.52 &6.61 \\
40 & 74 &21765  & 1.9 &2.0063(1) &1.0617(3) & 1.0314 &7.59 &8.04 \\
48 & 93 &21005  & 1.9 &2.0074(1) &1.0646(3) & 1.0342 &7.61 &7.83 \\
56 &136 &23795  & 1.6 &2.0084(1) &1.0674(2) & 1.0369 &7.62 &6.56 \\
64 &200 &26439  & 1.4 &2.0090(1) &1.0691(1) & 1.0385 &7.64 &7.18 \\
72 &150 &20000  & 1.8 &2.0093(1) &1.0701(2) & 1.0395 &7.65 &7.10 \\
80 &200 &25431  & 1.7 &2.00962(4)&1.0709(1) & 1.0403 &7.66 &6.77 \\
\hline
   \end{tabular}
  \end{center}
 \caption[a]{Measurement statistics at the simulation point $K_0 =
0.6929$:
$L$ is the
linear lattice size, $N_0$ is the number of cluster steps between
measurements, $N_{\rm
meas}$ stands for the number of measurements, $\tau_n$ is the
integrated
autocorrelation time of the charge density, $\langle e \rangle$ is
the
energy density, $\langle
n_{|q|} \rangle$ are the observed densities of dual cells with
charge $|q|$, and the total defect density
$\langle n \rangle$ is defined as
$\langle n \rangle = \langle n_1 \rangle + 4 \langle n_2 \rangle
+ 9 \langle n_3 \rangle + \dots $.}
\end{table}
%
%
%
\begin{table}              
\setlength{\tabcolsep}{0.80pc}
\newlength{\digitwidth} \settowidth{\digitwidth}{\rm 0}
\catcode`?=\active \def?{\kern\digitwidth}

 \begin{center}
  \begin{tabular}{|r|c|c|c|c|c|}
   \hline
\multicolumn{1}{|r|} {$L$}        &
\multicolumn{1}{c|}{$C$}      &
\multicolumn{1}{c|}{$C_q$}     &
\multicolumn{1}{c|}{$\chi_q$}    &
\multicolumn{1}{c|}{$\lambda_1$}  &
\multicolumn{1}{c|}{$\lambda_2$}  \\
\hline
 8 &2.177(18) &0.712(?6) &0.369(3) &2.423(19) &0.1225(?8) \\
12 &2.407(12) &0.833(?5) &0.430(2) &2.721(14) &0.1261(?6) \\
16 &2.562(27) &0.908(10) &0.467(4) &2.900(30) &0.1283(?9) \\
20 &2.651(36) &0.951(14) &0.486(6) &3.009(40) &0.1277(?7) \\
24 &2.752(40) &0.993(17) &0.506(8) &3.128(47) &0.1297(12) \\
32 &2.832(38) &1.027(14) &0.521(6) &3.223(43) &0.1308(12) \\
40 &2.970(45) &1.081(16) &0.542(7) &3.382(51) &0.1300(?9) \\
48 &2.977(44) &1.097(18) &0.553(8) &3.400(51) &0.1304(13) \\
56 &3.142(35) &1.157(12) &0.576(5) &3.587(38) &0.1309(13) \\
64 &3.181(30) &1.173(10) &0.579(5) &3.631(33) &0.1286(10) \\
72 &3.142(55) &1.155(19) &0.574(7) &3.585(62) &0.1315(14) \\
80 &3.182(49) &1.169(16) &0.579(6) &3.630(53) &0.1307(12) \\
\hline
   \end{tabular}
  \end{center}
 \caption[a]{Results for the matrix elements of the covariance
matrix $V \langle O_i;O_j \rangle$, with $O_1 = K e $,
$O_2 =  n $ at $K = 0.6930 (\approx K_c)$. Also included are
the eigenvalues $\lambda_1,\lambda_2$ of the covariance matrix.}
\end{table}
\clearpage
\newpage
%
  {\Large\bf Figure Headings}
  \vspace{1in}
  \begin{description}
    \item[\tt\bf Fig. 1:]
(a) $C_q = Td\langle n \rangle/dT$, and (b) the specific heat $C$
versus
$K$ for lattices of size $L = 12$, 40, 72. The
values were obtained by reweighting of the runs at $K_0 = 0.6929$.
    \item[\tt\bf Fig. 2:]
(a) $C_q$, (b) $C$, and (c) $\chi_q$ at $K = 0.6930 (\approx K_c)$
as function of the lattice size $L$.
The solid lines show
the best non-linear three-parameter fits to the data.
    \item[\tt\bf Fig. 3:]
    (a) $\langle n \rangle$, and (b) $\langle e \rangle$ at
$K = 0.6930 (\approx K_c)$ as
function of $L$, together with
the best non-linear three-parameter fits to the data.
    \item[\tt\bf Fig. 4:]
  (a) $\lambda_1$ and (b) $\lambda_2$ at $K = 0.6930 (\approx K_c)$
 as function of $L$.
Included are also
the best non-linear three-parameter fits to the data.

  \end{description}
\end{document}